# Non-Markovian dynamics for a two-atom-coupled system interacting with local reservoir at finite temperature

Li Jiang (姜丽)[1], Guo-Feng Zhang (张国锋)[1,2,3,4,*]

[1]*Key Laboratory of Micro-Nano Measurement-Manipulation and Physics (Ministry of Education), School of Physics and Nuclear Energy Engineering, Beihang University, Xueyuan Road No. 37, Beijing 100191, China*

[2]*State Key Laboratory of Software Development Environment, Beihang University, Xueyuan Road No. 37, Beijing 100191, China*

[3]*State Key Laboratory of Low-Dimensional Quantum Physics, Tsinghua University, Beijing 100084, China*

[4]*Key Laboratory of Quantum Information, University of Science and Technology of China, Chinese Academy of Sciences, Hefei 230026, China*

**Abstract:** By using the effective non-Markovian measure [H.P. Breuer, E.M. Laine, J. Piilo, Phys. Rev. Lett. **103**, 210401 (2009)], we investigate non-Markovian dynamics of a pair of two-level atoms (TLAs) system, each of which interacting with a local reservoir. We show that subsystem dynamics can be controlled by manipulating the coupling between TLAs, temperature and relaxation rate of the atoms. Moreover, the correlation between non-Markovianity of subsystem and entanglement between the subsystem and the structured bath is investigated, the results show that the emergence of non-Markovianity has a negative effect on the entanglement.

## 1. Introduction

A quantum dynamical semigroup usually is utilized to define the archetype of a Markovian process in an open quantum system, i.e., by the solutions of a master equation for the reduced density matrix with Lindblad structure [1,2]. However, in complex quantum systems one often encounters dynamical processes which sharply deviate from the relatively simple behavior predicted by a Markovian time evolution [3], in this case the assumption of a Markovian dynamics might be confronted with failure due to strong-coupling, finite-size environments or small time scales. Memory effects then become important, and the dynamics in this case is said to be non-Markovian.

The non-Markovianity was found to be usually associated with the occurrence of revivals, non-exponential relaxation, or negative decay rates in the dynamics. Recently, much effort has been devoted to the analysis of non-Markovian quantum evolution [4-16], many different measures of non-Markovianity have been proposed in the literatures to quantify memory effects in open systems, based on, for example, the quantum channels [17], the non-monotonic behaviors of

---

distinguishability [18-21], entanglement [22,23], Fisher information [24], correlation [25], the volume of states [26], capacity [27], the breakdown of divisibility [22,28], the negative fidelity difference [29], the non-zero quantum discord [30], the negative decay rates [31], and the notion of non-Markovian degree [32]. The criterion proposed by Breuer *et al.* [18] employs the trace distance between quantum states as a measure of non-Markovianity, which has been applied in experimental investigation of the non-Markovian behavior [33]. Even without knowing the properties of environment or the structure of the system-environment interaction, the scheme above is allowed to explore non-Markovianity experimentally.

In this paper, we consider a composite system: a two-qubit (depicted by two TLAs) system only coupled individually to their local thermal reservoirs besides the interaction between the two TLAs. One of the atom is assigned as the subsystem and the other atom as an auxiliary qubit, and then the subsystem we focus on can be recognized to be interacted with a structured bath (auxiliary atom + thermal reservoirs). How one could control the non-Markovian nature of the subsystem dynamics (single atom) by exploiting the features of being a part of the composite system is an intriguing issue. It is feasible to alter the properties of the subsystem dynamics, enabling one to induce a transition from Markovian to non-Markovian dynamics, by changing the atom-atom couplings, mean occupation number of the reservoir or the spontaneous emission rate of the atom. Moreover, we study how entanglement between the subsystem and the structured bath (auxiliary atom + thermal reservoirs) evolve at the critical border, where the subsystem undergoes a transition from Markovianity to non-Markovianity. In this sense, we show that mean occupation number of the reservoir has much stronger effect on entanglement than spontaneous emission rate of the atom in the critical region.

2. **The model and solutions**

We consider two TLAs **1** and **2** that present a two-qubit system and coupling with their local thermal reservoirs, besides direct interaction between the atoms. The Hamiltonian can be written as (we set $\hbar = 1$): [34]

$$H = \omega_0 \sum_{i=1}^{2} a_i^+ a_i + \frac{1}{2}v_0 \sum_{i=1}^{2} \sigma_i^z + g_0 \sum_{i=1}^{2}(a_i^+ \sigma_i^- + a_i \sigma_i^+) + \Omega(\sigma_1^+ \sigma_2^- + \sigma_1^- \sigma_2^+), \qquad (1)$$

where $\omega_0$, $v_0$, $g_0$ and $\Omega$ are constants and $\sigma^z$ denotes the usual diagonal Pauli matrix, $\frac{1}{2}v_0 \sum_{i=1}^{2} \sigma_i^z$ is the Hamiltonian of the two atoms $H_{at}$, the first term in right side of Eq. (1) is the two cavities (also considered as environments) $H_{cav}$ and the remaining are interactions (atom-cavity interaction and atom-atom interaction) $H_{int}$. The depiction of the time evolution of the qubits system is provided by the following master equation [2,35]

$$\frac{d\rho}{dt} = \frac{1}{2}(m+1)\gamma \sum_{i=1}^{2}\{[\sigma_i^-, \rho\sigma_i^+] + [\sigma_i^- \rho, \sigma_i^+]\}$$

$$+ \frac{1}{2}m\gamma \sum_{i=1}^{2}\{[\sigma_i^+, \rho\sigma_i^-] + [\sigma_i^+ \rho, \sigma_i^-]\} - i[\Omega(\sigma_1^+\sigma_2^- + \sigma_1^-\sigma_2^+), \rho]. \qquad (2)$$

where $\gamma$ ($\gamma > 0$) is the relaxation rate which is supposed to be the same for both qubits, and $\sigma_i^{\pm}$ ($i = 1; 2$) are the rasing (+) and lowering (−) operators of atom $i$, defined as $\sigma_i^+ = |1\rangle\langle 0|_i$, $\sigma_i^- = |0\rangle\langle 1|_i$, $m$ is the mean occupation number which also is assumed to be the same for both reservoirs and $\Omega$ is coupling strength between the two atoms. On the right hand side of equation (2), the first term describes the depopulation of the atoms due to simulated and spontaneous emission, while the second term corresponds to the re-excitations caused by the finite temperature.

The dynamics of the two-qubit ($S_1 S_2$) system would fall in the paradigm of quantum Markov processes [18,22] under the above circumstance. What we focus on is whether or not a subsystem (single qubit) dissipative dynamics could achieve transformation from Markovian to non-Markovian. The qubit $S_1$ can be treated as the subsystem and $S_2$ as the auxiliary qubit since the two qubits are identical according to Eq. (1) and $\zeta$ also can represents the environments. So it is deserved to draw more attention to the dissipative nature of the quantum dynamics of subsystem $S_1$. Generally, it is not easy to predict that qubit $S_1$ coupled to a structured bath (auxiliary qubit $S_2$ + environments $\zeta$), which represents an effective bath, would create peculiar characteristics of the dissipative process. In this sense, the variation of the properties of such structured bath, by changing the qubit-qubit coupling or the other parameters, would be meaningful for the current research.

We assume that our system is initially in an "X state" described by the following density matrix:

$$\rho(t) = \begin{pmatrix} a(t) & 0 & 0 & w(t) \\ 0 & b(t) & z(t) & 0 \\ 0 & z^*(t) & c(t) & 0 \\ w^*(t) & 0 & 0 & d(t) \end{pmatrix}. \quad (3)$$

Substituting (3) into (2), we obtain the following first-order coupled differential equation:

$$\dot{a}(t) = \gamma[-2(m+1)a(t) + mb(t) + mc(t)],$$
$$\dot{b}(t) = \gamma[(m+1)a(t) - (2m+1)b(t) + md(t)] + i\Omega[z(t) - z^*(t)],$$
$$\dot{c}(t) = \gamma[(m+1)a(t) - (2m+1)c(t) + md(t)] - i\Omega[z(t) - z^*(t)],$$
$$\dot{d}(t) = \gamma[(m+1)b(t) + (m+1)c(t) - 2md(t)],$$
$$\dot{z}(t) = \gamma[-(2m+1)z(t)] + i\Omega[b(t) - c(t)],$$
$$\dot{w}(t) = \gamma[-(2m+1)w(t)]. \quad (4)$$

These may be solved to yield the following expressions:

$$a(t) = \frac{1}{(2m+1)^2}\{m^2 + [2(a_0 - d_0)m^2 + (a_0 - d_0 + 1)m]X + [(2a_0 + 2d_0 - 1)m^2 + (3a_0 + d_0 - 1)m + a_0]X^2\},$$

$$b(t) = \frac{1}{(2m+1)^2}\{m(m+1) - [2(a_0 + 2c_0 + d_0 - 1 - (b_0 - c_0)\cos(\Omega t))m^2 + (a_0 + 4c_0 + 3d_0 - 2 - 2(b_0 - c_0)\cos(\Omega t))m + \left(c_0 + d_0 - 1 - \frac{1}{2}(b_0 - c_0)\cos(\Omega t)\right)]X - [(2a_0 + 2d_0 - 1)m^2 + (3a_0 + d_0 - 1)m + a_0]X^2\},$$

$$c(t) = \frac{1}{(2m+1)^2}\{m(m+1) + [2(a_0 + 2c_0 + d_0 - 1 - (b_0 - c_0)\cos(\Omega t))m^2 + (3a_0 + 4c_0 +$$
$$d_0 - 2 - 2(b_0 - c_0)\cos(\Omega t))m - \frac{1}{2}(b_0 - c_0)\cos(\Omega t)]X - [(2a_0 + 2d_0 - 1)m^2 + (3a_0 + d_0 -$$
$$1)m + a_0]X^2\},$$

$$d(t) = \frac{1}{(2m+1)^2}\{(m+1)^2 - (m+1)[2(a_0 - d_0)m + (a_0 + d_0 - 1)]X + [(2a_0 + 2d_0 -$$
$$1)m^2 + (3a_0 + d_0 - 1)m + a_0]X^2\},$$

$$z(t) = \left[z_0 + \frac{1}{2}i(b_0 - c_0)\sin(\Omega t)\right]X,$$

$$w(t) = w_0 X, \tag{5}$$

where $X = e^{-\gamma(1+2m)t}$, $a_0 = a(0)$, etc.

### 3. The measure of non-Markovianity

The trace distance between two quantum states $\rho(t)$ and $\tau(t)$ is a significant mean utilized for the measuremenf of the distinguishability of quantum states. A Markovian evolution, depicted by a dynamical semigroup of completely positive trace preserving (CPTP) maps whose nature induce the shrinkage of the trace distance between any fixed pair of initial states $\rho(0)$ and $\tau(0)$, can never augment the trace distance. An outflow of information from the system to the environment is a symbol of the reduction of trace distance which manifests the decrease of distinguishability between the two states. So a backflow of information into the system that interests us is interpreted by the contrary of shrinkage condition. In this scenario, a measure of non-Markovianity can be defined as in [18] by

$$N = \max_{\rho(0)\tau(0)} \int_{\sigma>0} dt\, \sigma(t, \rho(0), \tau(0)). \tag{6}$$

Here, $\sigma(t, \rho(0), \tau(0)) = \frac{d}{dt}D(\rho(t), \tau(t))$ is the rate of change of the trace distance, and

$$D(\rho(t), \tau(t)) = \frac{1}{2} Tr\, |\rho(t) - \tau(t)|, \tag{7}$$

where $|A| = \sqrt{AA^+}$ is the positive square root of $AA^+$ [36]. Therefore, the total increase of distinguishability during the total time evolution, i.e., the whole amount of information flowing back to the system which we are interested in, is represented by $N$. Under the circumstances, non-Markovian dynamics process would occur if and only if $N > 0$. In other words, an evolution is Markovian if and only if the trace distance of any two initial states decreases monotonically.

4. **Non-Markovian dynamics of subsystem**

In our case, we choose the system initial states $|10\rangle$ and $|00\rangle$, which means $b_0 = 1$ and $d_0 = 1$, otherwise $\rho_{i,j}(0) = 0$. Focusing on the single qubit reduced density matrix $\rho_{S_1}(t) = Tr_{S_2}[\rho_{S_1 S_2}]$, where $Tr_{S_2}[\rho_{S_1 S_2}]$ implies the trace of the auxiliary qubit degrees of freedom. So $\rho_{S_1}(t)$ can be written analytically as

$$\rho_{S_1}(t) = \begin{pmatrix} p_\pm & 0 \\ 0 & q_\pm \end{pmatrix}, \tag{8}$$

where $p_+$ and $q_+$, $p_-$ and $q_-$ represent the nonzero matrix elements of $\rho_{S_1}(t)$ for the initial states $|10\rangle$ and $|00\rangle$, respectively, which can be expressed as follows

$$p_+ = \frac{2m+[1+(1+2m)\cos(2\Omega t)]X}{2(1+2m)},$$

$$p_- = \frac{m-mX}{1+2m},$$

$$q_+ = 1 - p_+, q_- = 1 - p_-. \tag{9}$$

For the states Eq. (8), the trace distance can be obviously determined by

$$D\left(\rho_{S_1}(t), \rho'_{S_1}(t)\right) = \frac{1}{2}(|p_+ - p_-| + |q_+ - q_-|). \tag{10}$$

The trace distance $D\left(\rho_{S_1}(t), \rho'_{S_1}(t)\right)$ utilized to interpret the dissipative character of the subsystem dynamics. In Fig. 1, we show the trace distance between two quantum states of subsystem $S_1$ as a function of the qubit-qubit coupling strength $\Omega$ and time $t$ where the initial states of the two qubits are $|10\rangle$ and $|00\rangle$. As shown in Fig. 1, according to the criterion whether the trace distance for the single-qubit states is monotonic or not, the region where the red line locates may be a transition district from Markovian to non-Markovian.

Markovian process occurs when the trace distance for the single-qubit states is monotonic, in this case $\Omega$ is small as portrayed in the graph. As $\Omega$ increases, the trace distance for the single-qubit states becomes non-monotonic, which indicates that now there is a backflow of information from the structured bath (auxiliary qubit $S_2$ + environments ) to the subsystem $S_1$, and the dynamics of qubit $S_1$ is non-Markovian. So it is feasible to induce a transition from Markov to non-Markov behavior for the subsystem by changing the qubit-qubit coupling $\Omega$ for a fixed $\gamma$ and $m$.

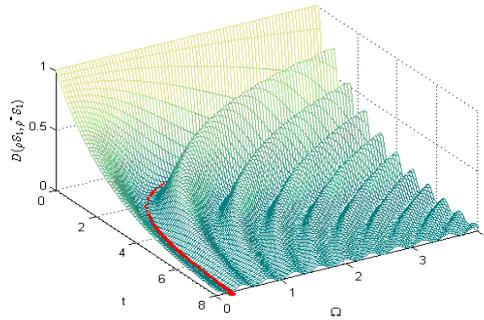

Fig. 1: The trace distance between two quantum states of the subsystem $S_1$ as a function of the qubit-qubit coupling strength $\Omega$ and time $t$, where the initial states of two qubits are $|10\rangle$ and $|00\rangle$. The other parameters $\gamma = 0.2$ and $m = 0.5$.

The effects of the parameter $\gamma$ and $m$ on the subsystem dynamical behaviors are another some intriguing problems, which is portrayed explicitly in Fig. 2 and Fig. 3. We can see that spontaneous emission rate of the atom $\gamma$ and the mean occupation number of the reservoir $m$ have the similar influence on the subsystem dynamics if the value of $\Omega$ is given. In the left plots of Fig. 2 and Fig.3, the qubit-qubit coupling $\Omega = 0.1$, and the trace distance for the single-qubit states is monotonic no matter what the variable value of $\gamma$ or $m$ is, which implies non-Markovian dynamics is not available for a weak coupling $\Omega$. While the right plots of Fig. 2 and Fig.3 shows that the qubit-qubit coupling $\Omega = 0.8$, the trace distance of the single-qubit reduced density matrix becomes non-monotonic with smaller value of $\gamma$ or $m$, which indicates that one can make a transition between the Markovian and non-Markovian dynamics for the subsystem by means of changing $m$ (or $\gamma$) when the qubit-qubit coupling $\Omega$ is large. Another interesting finding is that the time at which non-Markovian occurs is the same for a given $\Omega$ regardless of $\gamma$ and $m$.

From the discussion above, $\Omega$ plays a more important role in judging whether the subsystem undergoes non-Markovian dynamics or not.

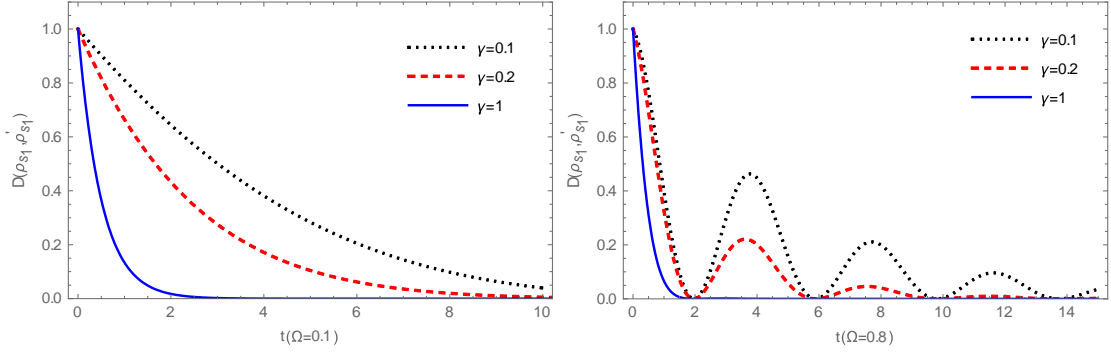

Fig. 2: The trace distance between two quantum states of the subsystem $S_1$ versus time $t$ with different spontaneous emission rate of the atom $\gamma$, where the initial states of two qubits are $|10\rangle$ and $|00\rangle$. The left plot: $\Omega = 0.1$ while the right plot: $\Omega = 0.8$. For both the plots $m = 0.5$.

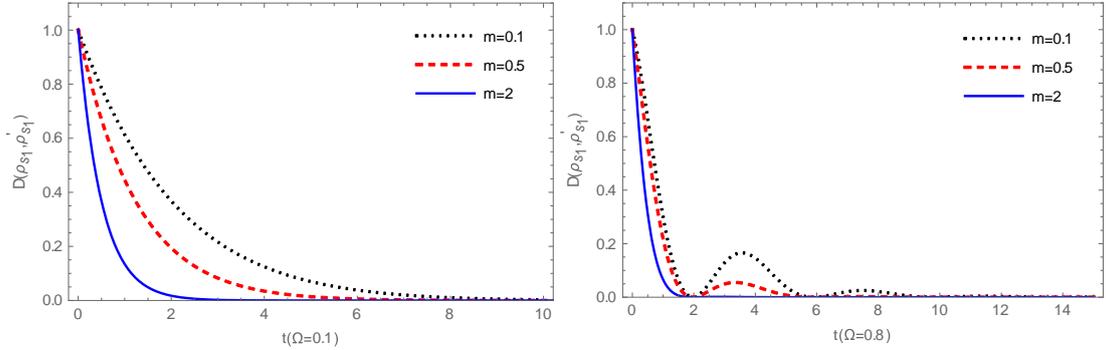

Fig. 3: The trace distance between two quantum states of the subsystem $S_1$ versus time $t$ with different mean occupation number of the reservoir $m$, where the initial states of two qubits are $|10\rangle$ and $|00\rangle$. The left plot: $\Omega = 0.1$ while the right plot: $\Omega = 0.8$. For both the plots $\gamma = 0.2$.

5.  **The entanglement and non-Markovianity**

Since the critical region between Markovian and non-Markovian regime is untraceable, what we want to explore is how the entanglement between the subsystem and the structured bath ($E_{S_1(S_2\zeta)}$) evolve **at the critical district**.

Mathematically, for a bipartite $SS'$ system, entanglement can be measured by its entropy of entanglement [36],

$$E(\rho_{SS'}) = S(\rho_S) = S(\rho_{S'}) , \qquad (11)$$

where, $S(\rho) = -Tr\rho \log_2 \rho$ is the von Neumann entropy. In our case, as mentioned above the initial state for two independent environments are vacuum state, we choose the two-qubit system being prepared initially in the separable initial state $|10\rangle$. Thus the whole '$SS'\zeta$ system' is described by an initial pure state ( $|\psi\rangle_{S_1 S_2 \zeta} = |10\rangle_{S_1 S_2} |00\rangle_{\zeta_1 \zeta_2}$ ), so we can calculate the entanglement between the subsystem $S_1$ and the structured bath ($S_2+\zeta$) directly from the entropy

$$E_{S_1(S_2\zeta)} = S(\rho_{S_1}) = S(\rho_{S_2\zeta}) , \qquad (12)$$

As we said before, we can obtain the analytical results of the nonzero matrix elements of $\rho_{S_1 S_2}$ for this initial state $|10\rangle$, then we get the matrix of the subsystem by tracing over the auxiliary qubit degrees of freedom $\rho_{S_1}(t) = Tr_{S_2}[\rho_{S_1 S_2}]$, hence we have

$$E_{S_1(S_2\zeta)} = -p_+ \log_2 p_+ - p_- \log_2 p_- \qquad (13)$$

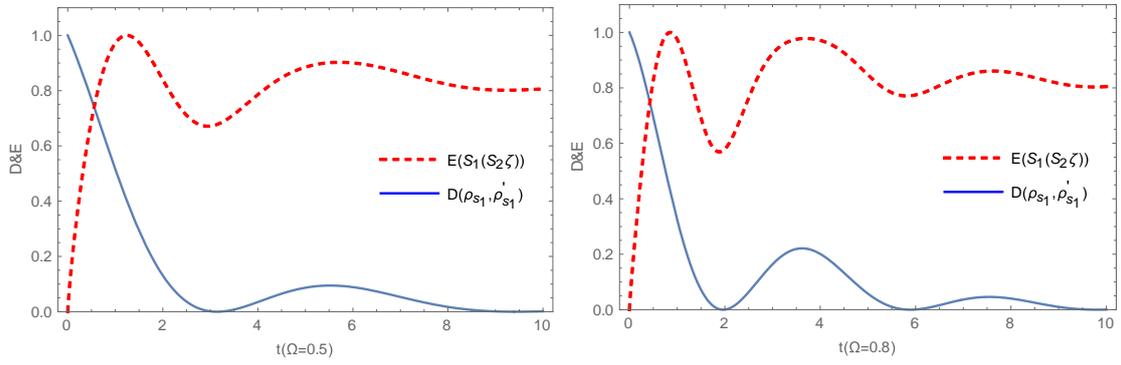

Fig. 4: Time evolution of the quantum entanglement between the subsystem $S_1$ and the structured bath $(S_2 + \zeta)$ and the trace distance between two quantum states of the subsystem $S_1$. The left plot : $\Omega = 0.5$, right plot: $\Omega = 0.8$. For both the plots $\gamma = 0.2$ and $m = 0.5$.

It can be seen in Fig. 4 that at $t = 0$, $E_{S_1(S_2\zeta)}$ equals zero because of the separability of the initial bipartite ($S_1$ and $S_2\zeta$) state. By comparing evolution of the trace distance $D(\rho_{S_1}, \rho'_{S_1})$ and the entanglement $E_{S_1(S_2\zeta)}$, it is exciting to see that $E_{S_1(S_2\zeta)}$ drops to minimal value when monotonicity of $D(\rho_{S_1}, \rho'_{S_1})$ becomes broken, which indicates the transition from Markovian to non-Markovian regime does harm to the dynamic process of the entanglement between the subsystem $S_1$ and the structured bath $(S_2 + \zeta)$. **In other words, if the quantum dynamics of the subsystem $S_1$ changes, there is an inevitable effect acted on the entanglement between the subsystem and the structured bath.**

Moreover, what we want to address now is that the effect of spontaneous emission rate of the atom $\gamma$ and mean occupation number of the reservoir $m$ on the entanglement between the subsystem and the structured bath **for the critical line**. In Fig. 5, we see that $E_{S_1(S_2\zeta)}$ achieves a steady value in the long-time limit and the steady value has close relation with the parameter $m$, while the other factor $\gamma$ only influences the time when approach the steady value rather than the magnitude of the steady value.

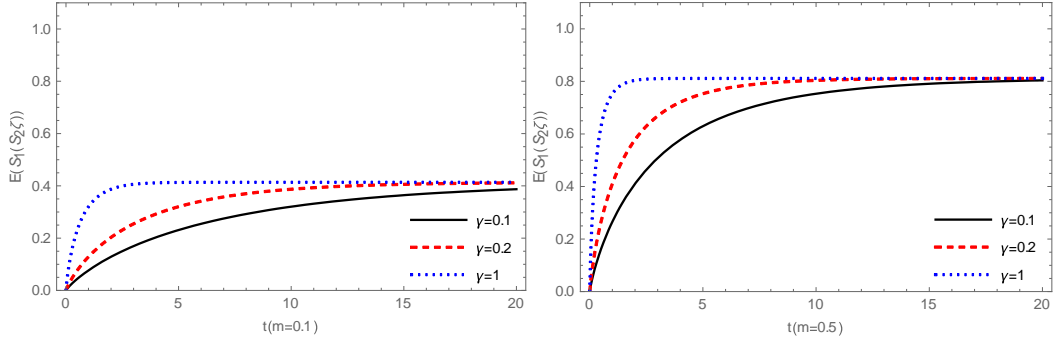

Fig. 5: The quantum entanglement between the subsystem $S_1$ and the structured bath $(S_2 + \zeta)$ versus time $t$ with different spontaneous emission rate of the atom $\gamma$ for the critical line, i.e. $\Omega t = \pi/2$ where a transition from Markovian to non-Markovian regime happens.

## 6.   Conclusions

In summary, we have proposed a composite system: a pair of two-level systems, each of which is interacting with a reservoir.   In this composite system, we have taken one qubit as the subsystem and the other qubit as an auxiliary qubit, and then the subsystem we have focused would be coupled to a structured bath (auxiliary qubit + environments). We have shown the feasibility of manipulating the non-Markovianity of the subsystem of interest, i.e., we have illustrated how the subsystem achieves a transition from Markovian to non-Markovian dynamics by changing the qubit-qubit coupling $\Omega$ or the parameter $\gamma$ and $m$, both $\gamma$ and $m$ make parallel yet limited contribution to the transition.

On the other hand, we have studied that the correlation between non-Markovian dynamics of the subsystem and the entanglement dynamics between the subsystem and the structured bath, and shown that the entanglement dynamics, which can be influenced in several different ways, counts on the mean occupation number of the reservoir m and spontaneous emission rate of the atom γ. It has been also shown that the emergence of non-Markovianity has the negative effect on the entanglement and finally the entanglement between the subsystem and the structured bath tends a steady value. In this sense, mean occupation number of the reservoir $m$ has the stronger effect rather than spontaneous emission rate of the atom $\gamma$ for obtaining a large-steady entanglement between the subsystem and the structured bath in our model.

**Acknowledgments**

This work is supported by the National Natural Science Foundation of China (Grant No. 11574022 and 11174024) and the Open Project Program of State Key Laboratory of Low-Dimensional Quantum Physics (Tsinghua University) grants Nos. KF201407.